**Evaluation of the surface strength of glass plates shaped by hot slumping process**

L. Proserpio[1], S. Basso[1], F. Borsa[1], O. Citterio[1], M. Civitani[1], M. Ghigo[1], G. Pareschi[1], B. Salmaso[1,2], G. Sironi[1], D. Spiga[1], G. Tagliaferri[1], A. D'Este[3], R. Dall'Igna[3], M. Silvestri[3], G. Parodi[4], F. Martelli[4], M. Bavdaz[5], E. Wille[5]

[1] INAF/Brera Astronomical Observatory, Via E. Bianchi 46, 23807 Merate (LC), Italy

[2] Università degli Studi dell'Insubria, Via Valleggio 11, 22100 Como (CO), Italy

[3] Stazione Sperimentale del Vetro -SSV-, Via Briati 10, 30141 Murano (VE), Italy

[4] BCV Progetti, Via S. Orsola 1, 20123 Milano (MI), Italy

[5] ESA European Space Agency, ESTEC, Keplerlaan 1, 2201 AZ Noordwijk, Netherlands

**Reference author:** Laura Proserpio, lproserpio@mpe.mpg.de, tel. +49-89-30000-3631

**Abstract:** The Hot Slumping Technology is under development by several research groups in the world for the realization of grazing-incidence segmented mirrors for X-ray astronomy, based on thin glass plates shaped over a mould at temperatures above the transformation point. The performed thermal cycle and related operations might have effects on the strength characteristics of the glass, with consequences on the structural design of the elemental optical modules and consecutively on the entire X-ray optic for large astronomical missions like for example IXO and ATHENA. The mechanical strength of glass plates after they underwent the slumping process was tested through destructive double-ring tests in the context of a study performed by the Astronomical Observatory of Brera with the collaboration of Stazione Sperimentale del Vetro and BCV Progetti. The entire study has been realized on more than 200 D263 Schott borosilicate glass specimens of dimension 100 mm x 100 mm and thickness 0.4 mm, either flat or bent at a Radius of Curvature of 1000 mm through the particular pressure assisted hot slumping process developed by INAF-OAB. The collected experimental data have been compared to non-linear FEM analyses and treated with Weibull statistic to assess the current IXO glass X-ray telescope design, in terms of survival probability, when subject to static and acoustic loads characteristic of the launch phase. The paper describes the activities performed and presents the obtained results.

# 1  Introduction

The Hot Slumping of thin glass foils is a very attractive technology under investigation by several groups for the realization of future segmented X-ray telescopes [1 - 5] aiming at combining a large effective area with good angular and energy resolution as for example the ones foreseen for IXO-like or ATHENA-like missions [6], [7]. Such telescopes assembling will be based on a principle of hierarchical integration: single mirrors segments will be integrated into elemental modules, usually called X-ray Optical Unit (XOU), then assembled and aligned in the Flight Mirror Assembly (FMA) through intermediate azimuthal structures to reestablish the cylindrical symmetry of the nested telescope with Wolter I (or a Wolter I approximation) optical design [8]. The slumping technology has already been successfully employed for the production of NuSTAR telescope, launched in 2012 [9], [10], able to deliver an image quality at focus of 45 arcsec HEW [11]. Additional studies are currently on going both by American and European research groups to further expand the technology in terms of achievable optical quality (the goal being 5 arcsec HEW) [12], [13]. The use of glass for the manufacturing of mirror segments, and possibly also for structural elements, ensures to meet the stringent mass requirements of space missions. However, this brittle material poses tight limits to the allowable stress level occurring during the entire life of the optical payload and requires a proper approach for the safety check during the design phases. Many different phenomena can concur to determine the stress levels during the different steps of payload manufacturing and mission operation, such as stresses induced during handling operation, transportation, ground testing, liftoff or thermal gradients, just to mention few examples. Moreover, the strength of glass is not an intrinsic property of the material rather it depends on the fabrication process and material history, the presence of flaws on the surface being the most relevant parameter, since flaws concentrate stress and reduce the theoretical strength. The distribution of such micro-cracks, the stress distribution and the size of the stressed area, the residual internal stresses from production process, the fracture toughness, and the "static fatigue" phenomenon are all elements to be taken into account when looking for strength parameter of glass components. The strength of glass component can therefore only be defined adopting a statistical approach based on experimental tests performed on specimens subjected to the same processes envisaged for the final parts since each phase of their life could in principle induce different critical defects.

The activities described in the present paper represent one of the first steps in the analysis of this complex argument. A similar study was internally conducted at NASA in the frame of Constellation-X mission [14]; compared to that, this work add a further steps in the analyses of results so to give the possibility of applying experimental data also to the effective stress distribution on the mirror modules and not only to the stress distribution recorded during tests. These activities have been realized in the frame of a study supported and coordinate by ESA and led by the Astronomical Observatory of Brera with the collaboration of other institutes like MPE and small enterprises like BCV-Progetti (Milano, Italy), ADS-International (Lecco, Italy) and Media Lario International (Bosisio Parini, LC, Italy) [15].

Even if the baseline for the realization of the IXO mirrors was represented by the so called "Silicon Pore Optics" technology investigated so far by ESA in collaboration with the Cosine company [16] and currently adopted for the ATHENA mirrors [17], the study is continuing with the scope of using the glass technology for the X-ray mirrors of other future missions. During the last 5 years, several prototypes have been realized [18], to show the potentiality of the INAF-OAB slumping and integration approach and a design of the complete IXO X-ray optical payload made by glass has been carried out [19]. One of the main scopes of this present study was to gain experimental number to demonstrate the goodness of the hypothesis on which this current design is based on.

The paper focalized on the evaluation of the surface strength of thin glass plates used for mirror segments shaped by hot slumping process: their strength has been characterized by the evaluation of the Weibull distribution, a classical statistic approach adopted to check the strength of brittle materials. The Weibull parameters have been evaluated by fitting experimental data coming from destructive double-ring tests on slumped plates and have been employed for the check of the current IXO telescope design. The paper is organized in 5 main sections according to the flow of activities sketched in figure 1.

## 2   SLUMPED SAMPLES REALIZATION

The statistical nature of the Weibull approach required the realization by thermal slumping of a large amount of glass samples, prepared following a process well representative of the final production of the X-ray segmented mirrors for the flight modules. The entire production chain comprises several steps, ranging from the procurement and selection of glass foils, their cutting to the required dimension for pursuing the hot shaping process, their cleaning, the realization of the thermal cycle, the post-cutting to the final dimension of the X-ray mirror segments, their coating with a reflective layer and the final integration into the XOU. If these steps are carried out in different locations, also packaging and shipment have to be considered. All these production steps could in principle affect the strength parameters of the final realized mirror. It is worth noting that further investigations are at present on going to improve the performances. The current procedures might therefore be subject to changes as result of the optimization and industrialization of the process. Since the current available laboratory set-up involves several steps still realized manually, with a lower degree of automation than the one expected to be available during mass production, the results obtained are representative of the current best knowledge of the process and should be considered conservative from the point of view of samples production.

*2.1 The hot slumping process with pressure assistance*

Different slumping processes exist: almost all share the basic idea of forming a thin glass mirror by shaping it over a mould through the application of a suitable thermal cycle that changes the viscosity properties of the glass allowing it to deform in order to assume a desired shape. Depending on the side of the forming mirror that comes in contact with the mould, two approaches are distinguished: the direct approach, in which the optical surface of the mirror comes in contact with the mould during the process, or the indirect approach, in which the contact happens on the back side [20]. The deformation of glass could take place only under its own weight (that is by gravity) or can be actively supported by the application of additional external forces. The particular process considered in the present work is known as *direct hot slumping with pressure assistance* [21], and is a direct approach characterized by the active application of pressure to help the glass reaching a full contact with the mould surface, ensuring the absence of mid-frequencies deformations (in the range between few mm and a couple of cm) that degrade the optical performances. During the course of the activities, lots of improvements have been obtained in the hot slumping technology with pressure assistance. In particular, a new method for pressure application has been developed: while the original approach developed at OAB makes use of a thin metal membrane for applying pressure on the glass plate being shaped so to force it in full contact with the mould, the new approach allows for the application of pressure directly on the glass plate, without intermediate materials [22]: in such a way, it is possible to avoid random local deformations and surface damages introduced by the metal pressing membrane that have detrimental effects both on mirror segments shape, meaning optical performances, and strength. In both cases, the entire process of slumping is realized inside a stainless steel muffle for thermal and cleaning reason: in the first case, the muffle is divided in two separate volumes by a 25μm-thick metal membrane, while in the second case the glass foil itself acts as a membrane to separate the muffle in two different chambers where a differential pressure can be established.

*2.2 Production of samples*

Accordingly, two main methods have been employed for the production of samples for the present study, i.e. the slumping with or without the use of metal pressing membrane (see figure 2 and 3). The rationale behind the decision of producing the specimens in the two ways was twofold. There was surely a temporal reason since at the beginning of activities, the new method for pressure application was not completely developed and the baseline was still considered the original one with the metal membrane. Furthermore, this decision gave the opportunity of performing a quantitative comparison of the two approaches with respect to the foil strength characteristic. In order to speed up the specimens' production, a stacking concept was initially considered: in every run, four glass foils were slumped together in stack, applying the pressure only on the last foil of the stack. To avoid their mutual sticking, sheets of the same material as the metal pressing membrane or Boron Nitride (BN) layers deposited on glass surface have been interposed between them, in the hypothesis that this situation was representative of the glass back surface condition, which during slumping comes in contact with the pressing membrane. However, after a number of tests, the envisaged solution came out to introduce spurious effects mainly related to cleaning issue, to the intrinsic structure of the thin membrane and to the dusty nature of BN layer. Meanwhile, the new pressure application method has been developed, and the stacking concept of foils within the muffle was no longer representative: in the new process in fact the back side of the glass foil being shaped is free, i.e. without contact with any external material. For this reason, the last samples have been produced by slumping each single glass plate at a time: this slow down the samples production and introduced a delay into the original agenda of activities but was fundamental to obtain reliable data.

In total, 42 slumping cycles have been carried out for the production of the more than 200 specimens. They all have been prepared using the thermal-pressure cycle reported in figure 4: after a first heating-up ramp at ~60°C/h, the maximum temperature of 570°C is keep for two hours before starting the cooling phases, which is divided in three steps: one at ~3°C/h, from $T_{max}$ up to $T_{annealing}$ (557°C) and, after a second plateau, the others at ~5°C/h and ~10°C/h. The one hour holding at annealing temperature guarantees the relaxation of major stresses inside the glass. After the controlled cooling phase, the oven is switch off and freely cools up to room temperature. A pressure of 50g/cm$^2$ is applied at reaching the highest cycle temperature and maintained till the opening of the oven.

The samples realization has been performed using the same materials as planned for IXO mirror manufacturing and following all the related phases (except for the reflective layer deposition and integration steps) as for the best current knowledge of the process. All the specimens are made of borosilicate glass type D263 of thickness 0.4 mm produced by Schott, slumped over a Zerodur K20 mould. These materials represent the base-line choice at the moment of writing: in particular, the glass type is already used for space application (e.g. NuSTAR) and is preferable with respect to other thin glass foils available on the market because of its relatively low characteristic temperatures, while the Zerodur K20 glass-ceramic material used for the mould offers the great advantage that it does not need any antisticking layer to prevent the adhesion of the D263 foils during the thermal cycles. One major drawback of this couple of materials is represented by their not-perfect matching of CTEs (Coefficient of Thermal Expansion), a problem that can be partially compensated for adopting the low cooling down rate. All specimens are slumped in bigger foils and then cut to the dimensions of 100 mm x 100 mm required for the strength tests. The cutting is performed with $CO_2$ laser technology, realized through the services of MDI Schott, Mainz - Germany. The majority of specimens is flat because of the better knowledge of double-ring test procedure and the consequent reliability in data analyses; some of them are cylindrical, with radius of curvature of 1000 mm, representative of an intermediate mirror segment in the current optical design of the IXO telescope.

*2.3 List of samples realized for the present study*

Table 1 reports a summary of all the realized specimens: they have been divided in different sets according to the specific procedure followed for their production and to the surface side that has been tested. In principle, the two surfaces of a slumped glass are different because of their different kind of contact during the shaping process: the optical surface in fact experiences contact with the polished surface of the mould, while the back surface comes in contact with the metal pressing membrane (for the original approach), or does not experience contact at all (in case of the new slumping approach). While the specimens produced with the original approach have been tested on both surfaces, specimens produced with the new approach (sets AIR-P and AIR-C) have been tested just on their back surface, since the side in contact with the mould should have in principle the same behavior as that obtained with the original slumping process. However, during tests realizations, a significant amount of failures have been registered also on the optical side, in spite of the lower tensile stress level induced by test set up. This happened because the contact of the glass foils with the mould at the optical side reduced somehow the glass strength with respect to the back side. For this reasons, depending on the side were the failure occurred, these specimens are identified with two set names and are considered representative respectively of the strength of the optical and of the non-optical side. A further set composed by not-slumped plates was also considered for comparison reasons: they were characterized in order to give a reference maximum value for glass strength that could be maintained after a slumping process. This value represents the best upper limit for untreated glass and is subject to unavoidable (though reducible with optimized operations) decrease during each phase of glass life.

As reported in table 1, the number of specimens for each set is variable and not uniform: this is mainly due to time constraint reasons and to a few breakages experienced during the cutting or shipment operations. On average, the amount of samples for each set is around 30 units. In spite of the fact that this number is quite small in order to derive statistical data, we believe that this preliminary analysis was very useful in allowing us to individuate weak points, fix some process procedures and develop experience to test slumped glass strength for implementing the obtained data in the structural design process. To our knowledge, it represents the only analysis of its kind performed so far for thin slumped foils to be used in space experiments.

## 3    MECHANICAL TESTS REALIZATION

The structural use of glass raises the issue of determining its failure strength. This implies reliable statistics regarding failure probability and the experimental determination of probability distribution parameters. The four-point bending test and ring-on-ring (RoR) test [23 - 26] are normally used for assessing whether glass plates complies with the pertinent product standard. A review of these test methods had been conducted in the past at SSV [37]; namely, the influence of the area under tensile stress and the effect of the non-uniform stress field were focused upon in order to estimate the material's Weibull parameters. Since no specific standard is available for glass plates 0.4 mm thick, existent testing method standards were taken as guideline only. Due to the broad strength data spread usually observed on brittle materials, EN 1288 standard prescribes using a large, but unspecified number of specimens when the testing purpose is to determine the characteristics or design bending strength of glass plates. As experimental data have to be statistically evaluated, the sample size for each tested set cannot be lower than thirty specimens [27].

Considering one testing method, we can say that the smaller the loaded area, the higher the average strength of the sample, because the probability of finding critical defects correspondingly decreases. Therefore the size of the loaded area is a fundamental element when evaluating experimental failure probability with different tests and geometries. Generally, using small loaded areas results in great strength value variability and it usually brings to overestimate the mechanical resistance of glass. Hence testing small specimens is recommended for comparison purpose only. Nevertheless, using physically based statistical fracture methods for the evaluation of breakage probability allows for the correlation of results obtained with similar stress fields, but on different areas [28], as long as the effective Weibull area doesn't prove to be less than 100 mm$^2$ [29]. Tests carried out using equal-sized loading areas subjected to different stress fields (e.g. uniaxial or biaxial tensile stress field), bring about different average strengths, i.e. higher failure stresses are observed under uniaxial stress fields. This is due to the crack's plane orientation relative to the principal stress directions: the crack-opening stress for a surface flaw not orthogonal to the uniaxial stress field is lower than the maximum principal stress, and the probability of unstable crack propagation is reduced. Therefore, failure probability increases for biaxial stress fields in which more crack orientations may prove to be critical.

Based on the previous observations, the importance of linking experimental strength data to the pertinent testing method and to the size of the loaded area becomes evident. Therefore, in order to evaluate the mechanical strength of the developed glass foil samples, ad-hoc ring-on-ring tests were carried out at Stazione Sperimentale del Vetro (SSV) laboratories in Murano (Venezia, Italy).

*3.1 Experimental set up*

During a coaxial double ring test the specimen rests on a bearing ring on the testing machine and an increasing load is applied perpendicularly to its upper surface by means of a second ring until the specimen fails. An electrostatic 3M polymeric film (approximately 100 µm thick) was applied on the upper surface of the specimens in order to keep together the glass fragments originated at failure, thus making the fracture analysis easier by allowing the experts to trace back the fracture origin (position where the fracture began) without affecting the test results. For flat specimens, a bearing toroid-ring with radius of 45 mm and a loading toroid-ring with radius of 30 mm (configuration R45-30, according to norm UNI-EN 1288:2001) were employed. For cylindrical specimens an ad-hoc double-ring test was developed. The curved specimens were tested with a modified R45-14 configuration (i.e. having a bearing ring with radius of 45 mm and a loading ring with radius of 14 mm) in order to generate a bi-axial stress field inside curved specimens while minimizing tensile stresses near the edges or constraints. The bearing ring was machined in order to follow the shape of the glass segment under testing and a perfect contact with the curved specimen was guaranteed, at least at the beginning of the test. The torus shape of the loading ring was not modified, but its radius was reduced so that the distance between the loading ring and the glass surface was lower than the thickness of the electrostatic film applied on the upper surface (i.e. facing-up surface, the one in contact with the loading ring) of the specimens. A specific metallic template was designed for the exact positioning of the curved specimens on the testing machine without interfering with the test implementation (see figure 5). Their concavity was placed upward, so that the nominal surface under test was the back one. All tests were performed in displacement control by means of an INSTRON-4411 dynamometer (max load 5 kN, resolution 0.1 N up to 400 N), with crossbar's displacement velocity of 1.39 mm/min (flat sample case) or 0.5 mm/min (curved sample case), corresponding to a stress increment in the samples of ~ 2 MPa/s.

*3.2 Data collection*

Load and displacement data were collected for each specimen. The slenderness of the specimens resulted in geometric non-linearity which prevents the application of analytical expressions. Consequently 3D non-linear FEM analyses of both testing apparatus (i.e. RoR for flat specimens or RoR for cylindrical specimens) were required for determining the actual stress field inside each specimen at breakage, and in particular the breakage stress at the failure origin. FEAs proved that the radial and circumferential (or tangential) stresses evocated in the specimen by the external load are not uniform nor equal to each other on the loading area (non-equibiaxial stress filed); moreover, high tensile stresses are also present on the upper surface of specimens. For flat specimens: on the facing-down surface the radial stresses are greater than tangential stresses and reach their maximum value at about 30 mm from the specimen's center, i.e. beneath the loading ring (see figure 6); on the facing-up surface the maximum tensile stress is reached by the radial stress at 45 mm from the center, in correspondence of the bearing ring, along the specimen's diagonal (see figure 7). In the same vein, for cylindrical specimens: on the facing-down surface, the radial tensile stresses are greater than the tangential ones (except in the center where their values are comparable) and increase from sample center up to the loading ring; on the facing-up surface, the maximum tensile stresses are in the radial direction, located at the bearing ring, nearby the sample diagonals.

To locate the failure origin, fractographic analyses were carried out on the specimens (except those failed at higher loads whose fragmentation prevented us to perform this analysis) and confirmed that the failures started where FEA identified the highest tensile stresses, as shown in figure 8 and 9. The fractographic analyses have been conducted by optical inspection of high resolution images taken by an OLYMPUS SZX12 stereo-microscope (up to 90X magnification). While for all the specimens of the sets TQ, STEEL, GLASS, and MOULD the fracture originated from the lower (facing-down) surface (as expected from test configuration), for the sets AIR-P and AIR-C the breakage started from the upper (facing-up) surface in 60 % and 35 % respectively (as shown in figure 10 and 11), due to the surface strength depletion caused by the mould.

For some specimens slumped in stack the fracture analyses showed defects on the glass surface which was in contact with other material during the forming process, i.e. the mould surface, the metal pressing membrane or one other glass foils in the stack (see figure 12). Typically, fractures originated from these defects ascribable to the slumping process and related phases (handling, cutting, cleaning). There was no evidence of such defects on the as-delivered specimens (set TQ). For sets AIR-P and AIR-C, the fracture analysis of specimens that had broken at the lower loads did not evidence the presence of particular surface defects close to the fracture origin (see figure 13 top). This demonstrates that the heavy defects observed on stack-slumped specimens were eliminated. Some secondary effects need to be better analyzed. One possible explanation is related to the crystalline structure of the mould material (Zerodur K20), characterized by the presence of crystal grain inside an amorphous matrix. These crystals causes an imprinting on the glass contact surface, both directly or because they make the cleaning harder. Better cleaning and glass strengthening methods should help in reducing the impact of these flaws: at the time of writing the use of lacquer to planarize the slumped glass surface is being analyzing [30]. On the back surface of some specimens of sets AIR-P$_{air}$ and AIR-C$_{air}$ a contamination of small metallic grains was found (see figure 13 bottom); its origin is due to the set up used for slumping that employed a stainless steel plate to shield the glass and mould from the direct heating of the electrical resistances in the oven, so to reduce at minimum thermal gradients. This element was positioned over the muffle at the very last step of the process preparation, right before closing the oven. A better control of the cleaning of this element will remove this contamination, which in any case does not represent a criticality.

## 4  GLASS WEIBULL PARAMETERS DETERMINATION

The mechanical resistance of glass is closely related to the characteristics of the flaws present on its surface (density, distribution and orientation of the superficial cracks). Thus, glass strength measurements are strongly dependent on the testing procedure through the actual size of the loading area (different areas have different probabilities of finding critical cracks) and the ensuing stress field (the probability that crack orientation will differ from principal stress directions). For a proper interpretation of glass fracture probability, physically-based statistical fracture theories were developed [31 - 37] that, starting from Weibull's *weakest link* theory [38], allow us to take into account the influence of the loading area extent and the actual stress field.

A statistical analysis of the failure probability data collected during the tests allowed evaluating the Weibull parameters that describe the mirror plates' strength. To evaluate fracture probability within a multi-axial stress field, the most simple fracture criterion was used: namely Fracture Mode I. For brittle materials like glass, it demonstrates a high level of concordance with experimental data [37]: once the existing surface flaws have been assimilated to flat cracks, the fracture expands when stress occurs in an orthogonal direction to a crack, surpassing the corresponding critical stress for Mode I, $\sigma_{Ic}$ (resistance of a crack placed orthogonally to uniaxial stress). According to Weibull's formulation, the average number of cracks in unit areas with mechanical resistance less than $\sigma_{Ic}$ can be expressed as:

$$N(\sigma_{Ic}) = \left(\frac{\sigma_{Ic}}{\sigma_0}\right)^{\beta} \tag{1}$$

The parameters **β** (module) and **σ₀** (reference resistance) depend on the fracture toughness of the material and statistical properties regarding the distribution of the crack's dimensions on the surface.

Assuming that the specimens are under plane stress, if we hypothesize that for surface cracks, all the directions contained in angle π have the same chance of being present, fracture probability for homogeneous defectiveness is derived from:

$$P = 1 - \exp\left[-\int_A \frac{1}{\pi}\int_\omega N(\sigma_{Ic} = \sigma_\perp)\,d\omega\,dA\right] \tag{2}$$

where: **A** is the superficial area under tensile stress; **σ⊥** is the tensile stress that's orthogonal to the crack in an arbitrary direction [39].

Expressing σ⊥ as a function of the principal stresses:

$$\sigma_\perp = \left[\sigma_1 \cdot \cos^2(\psi) + \sigma_2 \cdot \sin^2(\psi)\right] = \sigma_1 \cdot \left[\cos^2(\psi) + r_2 \cdot \sin^2(\psi)\right] \tag{3}$$

we can write (2) as follows:

$$P = 1 - \exp\left\{-\int_A\left[\frac{1}{\pi}\int_\psi\left(\frac{\sigma_\perp}{\sigma_0}\right)^\beta d\psi\right]dA\right\} = 1 - \exp\left\{-\int_A\left(\frac{C \cdot \sigma_1}{\sigma_0}\right)^\beta dA\right\} \tag{4}$$

where: **ψ** is the angle between the projection of the crack's normal on plane $\sigma_1$-$\sigma_2$ and the direction $\sigma_1$; **C** is the coefficient that takes into consideration the multi-axial nature of the actual stress field (C < 1 in case $\sigma_1 \neq \sigma_2$) calculated as follows [40, 41]:

$$C = \left\{\frac{2}{\pi} \cdot \int_0^\alpha\left[\cos^2(\psi) + \frac{\sigma_2}{\sigma_1} \cdot \sin^2(\psi)\right]^\beta d\psi\right\}^{1/\beta} \quad \text{with} \quad \begin{array}{ll} \alpha = \dfrac{\pi}{2} & \text{for } \dfrac{\sigma_2}{\sigma_1} \geq 0 \\[2mm] \alpha = \arctan\sqrt{-\left|\dfrac{\sigma_1}{\sigma_2}\right|} & \text{for } \dfrac{\sigma_2}{\sigma_1} < 0 \end{array} \tag{5}$$

Once the Weibull parameter **β** is known, the failure probability can be expressed as a function of the maximum stress reached on the surface of the specimen under testing ($\sigma_{max}$) taking into account the effective area ($S_{eff}$) which represents the superficial area that, if subjected to an uniform equibiaxial stress field equal to the maximum stress, would show the same failure probability as the actual stress field:

$$P = 1 - \exp\left\{-\int_A \left(\frac{C \cdot \sigma_1}{\sigma_0}\right)^\beta dA\right\} = 1 - \exp\left\{-\left(\frac{\sigma_{max}}{\sigma_0}\right)^\beta \cdot S_{eff}\right\} \tag{6}$$

$$S_{eff} = \frac{\int_A (C\,\sigma_1)^\beta dA}{\sigma_{max}^{\,\beta}} \tag{7}$$

The breakage load data and the ensuing principal stresses obtained from the pertinent numerical model were used to determine the Weibull parameters of the glass surface (module, **β**, and reference strength, **$\sigma_0$**) by best-fitting of the experimental failure probability of each sample set. For this purpose, an ad-hoc iterative algorithm based on the maximum likelihood method, as suggested by the ASTM C 1239-06A standard [42], was implemented in Matlab R2012b. Surface strength censoring caused by facing-up breakages in the AIR-P and AIR-C data sets, was also accounted for.

The step-by-step procedure for the iterative algorithm is as follow:

i) guess a value for the module, **β\*;**

ii) considering the maximum and intermediate principal stress fields evocated on the specimen subjected by each experimental breakage load, calculate the pertinent products $\left(\sigma_{max} \cdot S_{eff}^{1/\beta}\right)^*$;

iii) calculate the Weibull parameters (**β** , **$\sigma_0$**) using the maximum likelihood method on $\left(\sigma_{max} \cdot S_{eff}^{1/\beta}\right)^*$ data set;

iv) if $\left|\beta^* - \beta\right| > \varepsilon$, designate the module **β** obtained above as the new value of **β\***, and return to step ii) and repeat steps ii) and iii) until convergence is achieved.

Figure 14 shows the Weibull curve fitting for the MOULD set of experimental data of failure probability as a function of the maximum stress. The values of the Weibull module and characteristic parameter obtained by the analysis of the data (see [43], [44]) are summarized in table 2, together with the minimum breakage stress recorded for each sample. These minimum strength values cannot be directly applied to glass plates different in size or load configuration from those tested, since in glass objects the stress at failure is strictly related to the extension of the surface subject to tensile stress. On the contrary, the Weibull parameters can be used for the determination of the failure probability also for different geometries and loads combinations.

The experimental results show differences between untreated specimens (set TQ) and the slumped glass foils. This is immediately observable on figure 15, in which the fracture probability of the glass samples versus the maximum value of principal tensile stress is plotted. It clearly appears that the slumping process affected the characteristic strength of glass, lowering it. An improvement in the characteristic strength has been obtained with the new slumping approach. Indeed the failure probability curves for samples Air-P appear to be closer to that of the glass foils as delivered by the vendor (TQ), suggesting the possibility to maintain a good strength of the glass by controlling the glass-mould contact during the slumping process and all the related phases of handling and storage. The strength of the set Air-P$_{air}$ resulted to be higher than that of the untreated glass (lower failure probability); this contradicting result may be due to the relative low number of specimens on which the pertinent Weibull parameters relied upon: more accurate results can be obtained increasing the number of specimens. It's worth reminding that breakage stresses for flat and curved specimens are not directly comparable since they had been obtained from different testing set ups. In the experimental data range, the lower tail of the failure probability distribution proved to be overestimated. This effect, particularly evident for set STEEL and GLASS, is most likely due to the presence of a non-homogeneous defectiveness on the glass surface, as evidenced by the quite disperse stress values at breakage that justify the low values for the Weibull modulus (4 ÷ 4.5).

## 5  APPLICATION OF RESULTS TO IXO CASE

The first application of the results was the assessment of the current design of the IXO X-ray Telescope.

### 5.1  *The current IXO X-ray Telescope design based on slumped plates*

The major requirements for the IXO X-ray Telescope are a collecting area of around 2.5-3 $m^2$ @ 1.25 keV with 5" angular resolution. These can be fulfilled with a Wolter I telescope of 20 m focal length and comprising 350 mirror shells with radius of curvature ranging from 0.3 m to 1.7 m. The current design requires 16560 glass mirror segments, stacked into basic modules (X-ray optical units, XOU) through the use of glass spacers between segments of consecutive coaxial shells (ribs) and comprising two glass elements (back and front plane) giving stiffness to the entire structure [8]. 200 XOUs are arranged in 8 rings and 8 petals so to fill the available geometric area available of the telescope and re-establish the cylindrical symmetry of the nested Wolter I optical design. The current XOU configuration, reported in figure 16, is representative of an intermediate module belonging to ring 5 and composed by 40 plate pairs with an average radius of curvature of 1000 mm. The connecting ribs have the same length of the mirror foils, i.e. 200mm, and are bonded to the mirror foils along their whole length: they accomplish the twofold function of keeping the mirror segments in shape and in their mutual relative position while guaranteeing structural support to them.

The XOUs and IXO telescope design has been supported by a large set of FEA carried out on proper material model with the commercial software Ansys. Lacking a system level study specifically dedicated to the adoption of slumped glass technology for IXO units, reference mechanical and thermal environments have been retrieved from specifications relevant to Silicon Pore Optics technology, with some adjustment justified by the larger XOU mass when compared to a single SPO module [19].

In absence of any experimental data, an ultimate deterministic glass strength, in terms of maximum loads that can be sustained without breaking the mirrors, was stated and adopted during the design for of all glass components: allowable reference strength derived from experience and literature were assumed at the beginning equal to 6.7 MPa for long lasting loads and 10 MPa for short lasting and impulse loads. After this study, we can rely on experimental data to check the goodness of this hypothesis: the statistical strength distributions inferred from the experiment can be applied to relate the stress field that builds up inside the mirror segments (due to the load at which the optical payload is subject) with their survival probability.

The current analysis takes into account only one failure mechanism i.e. the mirror foils failure. Other glass elements (i.e. ribs and backplanes) have not been considered yet since available methods for strength improvements can be applied during their production, such us for example chemical etching or fire-polishing or tempering: all methods whose application to mirror segments is not straightforward and needs to be developed in order to preserve their precise optical shape.

*5.2 Assessment of the structural IXO Telescope design*

To check the goodness of the current IXO structural design the first step was to apply equation 4 to derive the survival probability of any single plate of any XOU when subject to the stress field generated by load conditions.

Simplifying hypotheses have been introduced for representing the loadings: only equivalent static loads and acoustic pressure at launch are considered since they represent the worst case for satellite payloads. In practice, the survival probabilities have been computed for the higher stress level (i.e. the one which gives the maximum failure probability) generated by the vectorial combination of equivalent static loads at launch (±70g longitudinal direction, 55g lateral direction) plus equivalent acoustic pressure (simulated by applying in radial direction a load equal to ±66g just on mirror foils and additional with respect to the quasi static loads). Also a bulk temperature variation at launch (equal to $\Delta T = \pm 20°C$, conservative case) has been assumed (see [46] for further details on loading conditions).

The Weibull parameters evaluated by SSV are relevant to the case of biaxial stresses, with equal principal tensile stresses. In this case the risk of failure is independent of flaw orientation, because a flaw of any orientation is exposed to the same stress. Instead the stress fields computed in the mirror foils by FEA are, obviously, unequal principal stresses and this reduces the risk of failure. This effect is taken into account by introducing a biaxial stress correction factor, according to the approach presented in [41]. This represents a step forward with respect to a previous work realized by NASA [14] in which no corrective factor was applied. The correction factor, as clearly visible on figure 17, depends on the ratio between the minimum and the maximum principal stress in each point ($\sigma_I$ and $\sigma_{II}$ being $\sigma_I$ the maximum tensile stress >0) and on the Weibull module $\beta$. It is computed in the hypothesis that all crack locations and orientations have the same probability of occurrence and that individual flaws do not influence each other. In addition, the evaluated parameters have been normalized taking into account the effective extension of the surface subject to tensile load (integration on foil surface A in the formula).

The survival probabilities related to each surface of the any mirror plates (i.e. front, back or edges) have been considered separately in order to discern the surface having the higher failure risk. To consider the effects of edges, the results of a previous tests campaign reported in table 2 have been employed [45]. The surface having the higher failure risk was then considered for the computation of the cumulative survival probability of a whole XOU composed by 40 plate pairs. The last step was the estimation of the survival probability of the whole FMA relying on the hypothesis that all the 200 XOUs present the same survival probability (in principle it could be different since the 8 different types of XOUs in the FMA could be subject to different input loads; however, as a first approximation, they are all considered identical). Effect of static fatigue has not taken into account in this first approach to the problem.

These steps have been performed for four cases, both with or without considering the effect of edges:

- Case 1 is representative of the best condition achievable in the unrealistic case where slumping process and related activities do not reduce at all mirror plates strength (use of Weibull parameters from set TQ).

- Case 2 is representative of the impact on glass plate's strength related to the initial slumping approach with metal membrane to apply pressure (use of Weibull parameters from set STEEL and MOULD).

- Case 3 is representative of the current new slumping process without the use of the metal pressing membrane, as evaluated with flat samples (use of Weibull parameters from set AIR-$P_{air}$ and AIR-$P_{mould}$).

- Case 4 is representative of the current new slumping process without the use of the metal pressing membrane, as derived from cylindrical samples (use of Weibull parameters from set AIR-$C_{air}$ and AIR-$C_{mould}$).

A major important hypothesis assumed in the analyses concerns the definition of a catastrophic failure. The evaluation of the consequences in case one or several mirror plates break is not a trivial task and it would require engineering activities at the system level. So at the moment we rely on the very severe assumption that the breakage of just one out of the 16560 mirror plates which compose the whole FMA determines a catastrophic failure, i.e.

could lead to the complete mission failure. In other words, in order to avoid catastrophic breakage, it has been required that no one among the 16560 mirror foils of the FMA crashes, which can probably be considered a worst case scenario.

The results suggest that the glass foil material in condition "as delivered" guarantee a survival probability at FMA level of 99.999% when the possibility of failure at the edges is excluded, and of 99.23% when failure at the edges is taken into account. A relevant worsening of the strength, both on optical and back mirror surfaces, is recorded after the application of the "old standard slumping process": in this case in fact, failure probability is increased by a factor of 12000 at the optical surface and of 39000 at the back surface, with a drop in the survival probability of FMA to 79.60% or 80.22% if failure at edges are excluded. The general worsening masks the impact of the breakage at the edges. The situation is greatly better when considering the "new slumping process". The failure probability at the optical front surface is almost confirmed but the back surface strength gains two order of magnitude with respect to the previous case, resulting in a survival probability at FMA level equals to 96.27%, or 97.02% if failures at edges are excluded. The result is confirmed as for flat and cylindrical samples, with only statistical minor differences. The results are promising and suggest that, under the considered hypotheses, the level of 99% survival probability of the FMA mirror plates seems achievable with the present XOU design, with only minor optimization on the slumping technique and an improvement of the cutting phase. Furthermore, in order to increase the survival probability of the mirror segments assembly, it should be considered the possibility to realize a procedure by scanning and/or by load tests for a preliminary proof test to be applied to the mirror plates before integration, in order to identify the weak plates to be discarded.

We are aware of the limitations of the presented analysis: the reliability of the Weibull parameters describing the mirror plates strength is affected by the relatively scarce number of specimens available and the statistical strength distributions suffer from the smallness of tested area, in comparison to the total surface of the Flight Mirror Assembly (FMA) mirror plates: at the moment, each specimen allows to asses few square centimeters of surface while the amount of glass foils in the FMA reaches much larger areas (around 1200m$^2$). The adopted two-parameter Weibull approach assumes that all stress levels, even very low, give a contribution to the failure probability. So, even with very low stresses, for large stressed area as in our case there is a not-negligible failure probability. At this regard, this approach could become too severe and we think a three-parameter Weibull function should be adopted for future activities. Furthermore, the Weibull distribution is applied outside the range of the experimental values, on the lower tail of the probability distribution, in a zone where no experimental evidence exists about their representativeness of the real statistical strength distribution. The feeling is that the distribution obtained could be particularly severe, when extrapolated to the stress level envisaged in the mirror plates (see [47]). This feeling is also supported by simple evaluations based on linear fracture mechanic. The flaws size at failure has been evaluated as a function of the tensile stress acting in direction normal to surface defect, according to linear fracture mechanic criteria and assuming for the mirror foil material a fracture toughness $K_{IC}$ = 0.75 MPa·m$^{1/2}$ [41]. It comes out that, at the stress level assumed for the X-ray Optical Unit (XOU) design, quite large cracks are necessary to trigger the glass failure. In case of through thickness-cracks the failure flaw size is in the range 2.5-3mm; while for "thumbnail" cracks, stress considerably larger (at least a multiplying factor 2.5-3) respect those assumed in design phases are necessary, in order to activate the failure, starting in any case from initial cracks 1-4mm long and 0.25t-0.6t deep (t=mirror foil thickness).

Despite these limitations, this preliminary approach to the problem represents an important step forward in the comprehension of the slumping technique and gave us the possibility to look deeply inside some phases of the slumping process and related activities pointing out the direction for optimization.

# 6  CONCLUSION

The surface strength of glass after the application of a slumping thermal-pressure cycle has been evaluated through double ring destructive tests. More than 200 specimens have been tested to take into account the effect of the "old slumping process" (in which the pressure was applied through metal membrane) or of the "new slumping process" (in which pressure is directly applied on the glass foil). All samples have been realized following all the steps (but coating deposition and integration) envisaged for the final mirror segments, at the best current knowledge of the process. In particular, 0.4 mm thick D263 glasses have been slumped on Zerodur K20 and cut after that at dimension of 100 mm x 100 mm, as required for the double-ring tests. The empirical values recorded have been compared with non-linear Finite Element Analyses to derive the stress field in the glasses at breaking. A statistical analysis allowed for the determination of Weibull parameters of glass, which gives the failure probability associate to a particular stress value. The first application of the derived parameters consisted on a reliability assessment of the current IXO structural design. Present results are promising suggesting that the level of 99% survival probability of the mirror foils of the whole FMA under equivalent static load seems reachable with the present XOU design, considering only minor improvement in the slumping technique and an optimization of the cutting technology, despite the strength reduction of glass after slumping, mainly coming from surface damages due to the contact between glass and other materials employed during the mirrors manufacturing.

*AKNOWLWDGEMENT*

This research was supported by ESA, in the frame of CCN1 of contract # 22545.

*REFERENCES*

1. Charles J. Hailey, Finn E. Christensen, William W. Craig, Fiona A. Harrison, Jason E. Koglin, Robert Petre, Haitao Yu, and William W. Zhang, "Fabrication and performance of Constellation-X hard x-ray telescope prototype optics using segmented glass," Proc. of SPIE Vol. 5168, (2004)

2. Mauro Ghigo, Oberto Citterio, Francesco Mazzoleni, Giovanni Pareschi, Bernd Aschenbach, Heinrich Braeuninger, Peter Friedrich, Guenter Hasinger, Thorsten Doehring, Hauke Esemann, Ralf Jedamzik, Eva Hoelzel and Giancarlo Parodi, "The manufacturing of the XEUS x-ray glass segmented mirrors: status of the investigation and last results," Proc. of SPIE Vol. 5168, (2004)

3. Ghigo, M., R. Canestrari, L. Proserpio, E. Dell'Orto, S. Basso, O. Citterio, Pareschi, G., and G. Parodi, "Slumped glass option for making XEUS mirrors: preliminary design and ongoing developments," Proc. of SPIE Vol. 7011, (2008)

4. M. Skulinova, R. Hudec, J. Sik, M. Lorenc, L. Pina, and V. Semencova, "New Light Weight X-Ray Optics - Alternative Materials," Proc. of SPIE Vol. 7360, (2009)

5. Monika Vongehr, Peter Friedrich, Peter Predehl, and Heinrich Bräuninger, "Development of slumped glass mirror segments for large x-ray telescopes," Proc. of SPIE Vol. 7011, (2008)

6. Jay Bookbinder, "An overview of the IXO Observatory," Proc. of SPIE Vol. 7732, (2010)

7. N. Rando, D. Martin, D. Lumb, P. Verhoeve, T. Oosterbroek, M. Bavdaz, S. Fransen, M. Linder, R. Peyrou-Lauga, T. Voirin, M. Braghin, S. Mangunsong, M. van Pelt, and E. Wille, "Status of the ESA L1 mission candidate ATHENA," Proc. SPIE. Vol. 8443, (2012)

8. M. Civitani, S. Basso, M. Bavdaz, O. Citterio, P. Conconi, D.Gallieni, M. Ghigo, B. Guldimann, F.Martelli, G.Pagano, G. Pareschi, G. Parodi, L. Proserpio, B.Salmaso, D. Spiga, G. Tagliaferri, M.Tintori, E.Wille, and A. Zambra, "IXO X-Ray mirrors based on slumped glass segments with reinforcing ribs: optical and mechanical design, image error budget and optics unit integration process," Proc. of SPIE Vol. 7732, (2010)

9. William W. Craig, HongJun An, Kenneth L. Blaedel, Finn E. Christensen, Todd A. Decker, Anne Fabricant, Jeff Gum, Charles J. Hailey, Layton Hale, Carsten B. Jensen, Jason E. Koglin, Kaya Mori, Melanie Nynka, Michael J. Pivovaroff, Marton V Sharpe, Marcela Stern, Gordon Tajiri, and William W. Zhang, "Fabrication of the NuSTAR Flight Optics," Proc. of SPIE Vol. 8147, 81470H (2011)

10. Fiona A. Harrison et al, "THE NUCLEAR SPECTROSCOPIC TELESCOPE ARRAY (NuSTAR) HIGH-ENERGY X-RAY MISSION," The Astrophysical Journal, 770:103 (19pp), 2013 June 20

11. Nicolai F. Brejnholt et al., "The Rainwater Memorial Calibration Facility for X-Ray Optics," Hindawi Publishing Corporation X-Ray Optics and Instrumentation Volume 2011, Article ID 285079, 9 pages doi:10.1155/2011/285079 (2011)

12. W.W. Zhang, M. Atanassova, M. Biskach, et al., "Mirror Technology Development for the International X-ray Observatory Mission (IXO)," Proc. of SPIE Vol. 7732, (2010)

13. M. Ghigo, S. Basso, F. Borsa, O. Citterio, M. Civitani, P. Conconi, G. Pagano, G. Pareschi, L. Proserpio, B.Salmaso, G. Sironi, D. Spiga, G. Tagliaferri, A. Zambra, G. Parodi, F. Martelli, D. Gallieni, M. Tintori, M. Bavdaz, E. Wille, "Development of high angular resolution x-ray telescopes based on slumped glass foils," Proc. SPIE 8443, (2012)

14. C. He, BATC et al., "Constellation-X: Glass Strength Investigation for SXT, Constellation-X: BATC," Memorandum No. MEB-2008-010 dated 3-25-2008 (2008)

15. G. Pareschi, S. Basso, M. Bavdaz, O. Citterio, M. M. Civitani, P. Conconi, D. Gallieni, M. Ghigo, F. Martelli, G. Parodi, L. Proserpio, G. Sironi, D. Spiga, G. Tagliaferri, M. Tintori, E. Wille and A. Zambra, "IXO glass mirrors development in Europe," Proc. of SPIE Vol. 8147, 81470L (2011)

16. Maximilien J. Collon, Ramses Günther, Marcelo Ackermann, Rakesh Partapsing, Giuseppe Vacanti, Marco W. Beijersbergen, Marcos Bavdaz, Kotska Wallace, Erik Wille, Mark Olde Riekerink, Jeroen Haneveld, Arenda Koelewijn, Coen van Baren, Peter Müller, Michael Krumrey, Michael Freyberg, Anders C. Jakobsen, and Finn Christensen, "Design, fabrication, and characterization of silicon pore optics for ATHENA/IXO," Proc. of SPIE Vol. 8147, 81470D. (2011)

17. ATHENA Assessment Study Report (Yellow Book), 2012. ESA website

18. Marta Maria Civitani, Stefano Basso, Oberto Citterio, Paolo Conconi, Mauro Ghigo, Giovanni Pareschi, Laura Proserpio, Bianca Salmaso, Giorgia Sironi, Daniele Spiga, Gianpiero Tagliaferri, Alberto Zambra, Francesco Martelli, Giancarlo Parodi, Pierluigi Fumi, Daniele Gallieni, Matteo Tintori, Marcos Bavdaz, and Eric Wille, "Accurate integration of segmented x-ray optics using interfacing ribs," *Opt. Eng*. 52(9), 091809 (Jun 24, 2013). doi:10.1117/1.OE.52.9.091809

19. G. Parodi, F. Martelli, S. Basso, O. Citterio, M. Civitani, P. Conconi, M. Ghigo, G. Pareschi, and A. Zambra, "Design of the IXO optics based on thin glass plates connected by reinforcing ribs," Proc. of SPIE Vol. 8147, (2011)

20. A. Winter, M. Vongehr, and P. Friedrich, "Light weight optics made by glass thermal forming for future X-ray telescopes," Proc. of SPIE Vol. 7732, (2010)

21. L. Proserpio, M. Ghigo, S. Basso, P. Conconi, O.Citterio, M. Civitani, R. Negri, G. Pagano, G. Pareschi, B. Salmaso, D. Spiga, G. Tagliaferri, L. Terzi, A. Zambra, G. Parodi, F. Martelli, M. Bavdaz, and E. Wille, "Production of the IXO glass segmented mirrors by hot slumping with pressure assistance: tests and results," Proc. of SPIE Vol.8147, (2011)

22. Mauro Ghigo, Laura Proserpio, Stefano Basso, Oberto Citterio, Marta M. Civitani, Giovanni Pareschi, Bianca Salmaso, Giorgia Sironi, Daniele Spiga, Giampiero Tagliaferri, Gabriele Vecchi, Alberto Zambra, Giancarlo Parodi, Francesco Martelli, Daniele Gallieni, Matteo Tintori, Marcos Bavdaz, Eric Wille, Ivan Ferrario, and Vadim Burwitz, "Slumping technique for the manufacturing of a representative x-ray grazing incidence mirror module for future space missions," Proc. of SPIE Vol. Optifab 8884, (2013)

23. EN 1288-1:2000, Glass in building – "Determination of the bending strength of glass - Part 1: Fundamentals of testing glass", *CEN* (2000).

24. EN 1288-2:2000, Glass in building – "Determination of the bending strength of glass - Part 2: Coaxial double ring test on flat specimen with large test surface area", *CEN* (2000).

25. EN 1288-3:2000, Glass in building – "Determination of the bending strength of glass - Part 3: Test with specimen supported at two points (four-point bending)", *CEN* (2000).

26. EN 1288-5:2000, Glass in building – "Determination of the bending strength of glass - Part 5: Coaxial double ring test on flat specimen with small test surface area", *CEN* (2000).

27. ASTM C 158-02, "Standard tests methods for strength of glass by flexure (Determination of modulus of rupture)", *American society for Testing Materials* (2008).

28. M. H. Krohn, J. R. Hellmann, D. L. Shelleman, C. G. Pantano, G. E. Sakoske, "Strength and Fatigue of float glass before and after enameling", *The Glass Researcher*, **11**, 2, 24-28 (2002).

29. A. A. Wereszczak, K. K. Kirkland, M. E.Ragan, K.T. Jr. Strong, H. Lin, "Size scaling of tensile failure stress in float soda-lime-silicate glass", *Int. J. App. Glass Sci.*, **1**, 2, 143-150 (2010).

30. B. Salmaso, A. Bianco, O. Citterio, G. Pareschi, G. Pariani, L. Proserpio, D. Spiga, D. Mandelli, and M. Negri, "Micro-roughness improvement of slumped glass foils for x-ray telescopes via dip coating," Proc. of SPIE Vol. 8861, (2013)

31. A. G. Evans, "A general approach for the statistical analysis of multiaxial fracture", *Journal of the American Ceramic Society*, Volume **61**,7-8, 302 – 308 (1978).

32. S. B. Batdorf, H. L. Heinisch Jr., "Weakest link theory reformulated for arbitrary fracture criterion",

*Journal of the American Ceramic Society*, **61**,7-8, 355 – 358 (1978).

33. S. B. Batdorf, G. Sines, "Combining data for improved Weibull parameter estimation", *Journal of the American Ceramic Society*, **63**, 3-4, 214–218 (1980).

34. D. K. Shetty, A. R. Rosenfield, W. H. Duckworth, P. R. Held, "A Biaxial-flexure test for evaluating ceramic strengths", *Journal of the American Ceramic Society*, **66,** 1, 36–42 (1983).

35. L. Y. Chao, D. K. Shetty, "Equivalence of physically based statistical fracture theories for reliability analysis of structural ceramics in multiaxial loading", *Journal of the American Ceramic Society*, **73,** 7, 1917–1921 (1990).

36. A. Bruckner-Foit, T. Fett, K. S. Schirmer, D. Munz, "Discrimination of multiaxiality criteria using brittle fracture loci", *Journal of European Ceramic Society,* **16**, 1201-1207 (1996).

37. R.Dall'Igna, A.D'Este, M.Silvestri, *Comments on test methods for the determination of structural glass strength*, ATIV XXV Conference 18-19 Nov. 2010

38. W. Weibull, "A statistical distribution function of wide applicability", *Journal of Applied Mechanics*, **18**, 293-297 (1951).

39. Munz D., Fett T., Ceramics. Mechanical properties, failure behaviour, materials selection, Springer, (1999) 181-189

40. W. L. Beason, "A failure prediction model for window glass", *Institute for Disaster Research, Texas Tech University Press* (1980).

41. W. L. Beason, and J. R. Morgan, "Glass failure prediction model", *Journal of Structural Eng.*, **110**, 2, 197–212 (1984).

42. ASTM C 1239-06A, "Standard practice for reporting uniaxial strength data and estimating Weibull distribution parameters for advanced ceramics"*, American society for Testing Materials* (2004).

43. Glass in building. Procedures for goodness of fit and confidence interval for Weibull distributed glass strength data, EN 12603:2002

44. Standard Practice for Reporting Uniaxial Strength Data and Estimating Weibull Distribution Parameters for Advanced Ceramics, ASTM C1239-06A

45. Internal document: Resistenza meccanica del bordo di lastre di vetro boro-silicato, Report SSV N.92372 – 23/12/2009

46. ESA contract 22545 document: IXO PROJECT - Design of a X-ray Optical Unit using glass mirrors, IXO-BCV-RE-002 – Issue 1- 29/9/2011

47. Internal document: Resistenza meccanica di lastre Schott D263 lavorate a caldo, Report SSV N.108932 –7/5/2013

**Biographies**

Biographies of the authors are not available.

**Tables**

**Table 1:** Summary of the realized and tested samples: they were all made of glass type D263 slumped on Zerodur K20 mould and had dimension of 100 mm x 100 mm, with thickness of 0.4 mm. All samples were $CO_2$ laser cut at their edges after slumping.

| SET | # | SHAPE | SLUMPING REALIZATION | TESTED SURFACE[*] | Notes |
|---|---|---|---|---|---|
| TQ | 49 | flat | n.a. | Both[**] | Samples as delivered by vendor, give the maximum theoretical strength for untreated glass. |
| STEEL | 31 | flat | Old approach, in stack with interposed metal membrane | Back | Tested on back surface that has lower strength because of its contact during slumping with metal pressing membrane or BN layer. Presence of imprinting of the micro-structure of the anti-sticking interlayers. |
| GLASS | 45 | flat | Old approach, in stack with interposed BN layer | Back | |
| $MOULD_1$ | 16 | flat | Old approach, stack configuration | Optical | Obtained by the same experimental tests of set STEEL and GLASS, considering the lower glass in the stack that during slumping was in contact with the mould. |
| $MOULD_2$ | 18 | flat | | Optical | |
| AIR-$P_{air}$ | 28 | flat | New approach: Pressure directly applied on the free-back surface of glass | Back | All tested on the back surface; however, depending on the side where the failure occurred, they are considered separately representative of the optical or back surface. |
| AIR-$P_{mould}$ | 30 | flat | | Optical | |
| AIR-$C_{air}$ | 32 | cyl. | | Back | |
| AIR-$C_{mould}$ | 17 | cyl. | | Optical | |
| TOT. | 266 | | | | |

[*]*Back surface* means the surface of the glass that during the slumping process did not come in contact with the mould, and were in contact with metal membrane or Boron Nitride interposed antisticking layer for STEEL and GLASS or was not in contact with anything for set AIR-P and AIR-C.
[**]*Both* apply in the considered hypothesis that the surfaces of the glass are exactly the same when it is delivered by the vendor given the symmetry of the down-drawn production process.

**Table 2:** Weibull Module (ß) and Weibull characteristic parameter ($\sigma_0$) for Schott D263 glass plates examined in the present work. The minimum tensile stress $\sigma_{min}$ at breaking point is also reported.

| Specimens set | ß | $\sigma_0$ [MPa mm$^{2/\text{ß}}$] | $\sigma_{min}$ [MPa] |
|---|---|---|---|
| TQ | 6.7 | 597.9 | 117 |
| STEEL | 4.3 | 612.8 | 43 |
| GLASS | 4.6 | 595.0 | 53 |
| MOULD | 4.8 | 488.1 | 76 |
| AIR-P$_{air}$ | 5.0 | 926.4 | 90 |
| AIR-P$_{mould}$ | 4.7 | 597.5 | 74 |
| AIR-C$_{air}$ | 5.7 | 419.9 | 110 |
| AIR-C$_{mould}$ | 6.3 | 241.2 | 82 |
| *Laser cut | 3.39 | 674.1 | 76 |

*SET composed by 50 samples previously tested to check laser cutting effects on glass edge strength [45] and whose results have been included in the following.

**List of figure**

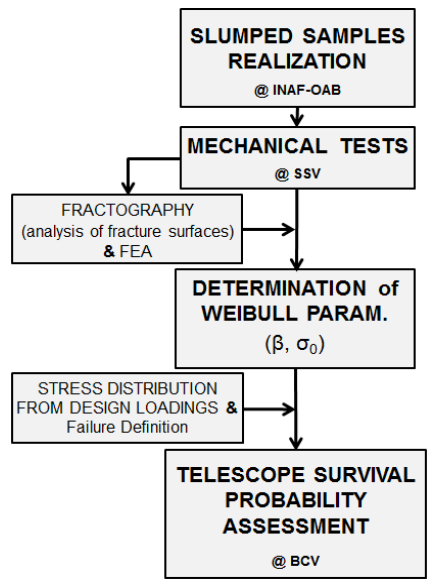

**Fig. 1:** Flow of the activities presented in the paper.

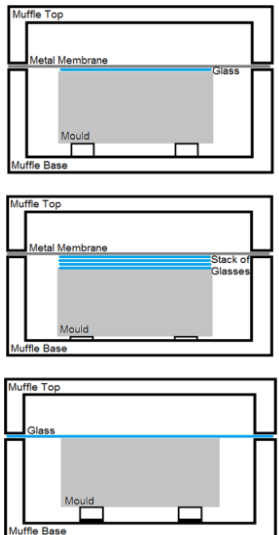

**Fig. 2:** Schematics of the slumping approaches followed for the production of samples considered in this study: (Top) pressure application through metal membrane; (Center) stacking concept to speed up the production by slumping several glass plates at a time. Between the glass foils in the stack, metallic or BN layer has been used as antisticking (not shown in the schematic picture); (Bottom) pressure application directly on the glass plate, without any intermediate material.

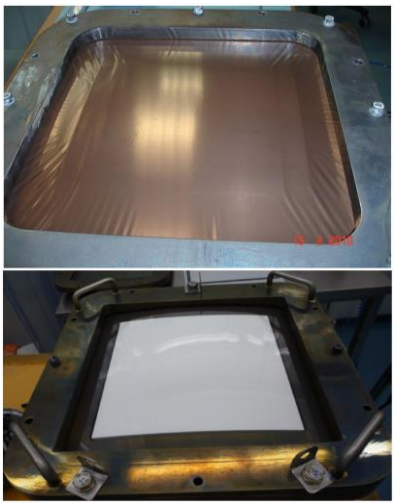

**Fig. 3:** Comparison between the two INAF-OAB approaches for pressure application during slumping. (Top): the application of pressure is realized through the use of metallic membrane. (Bottom): the glass foil itself act as a membrane allowing for pressure application.

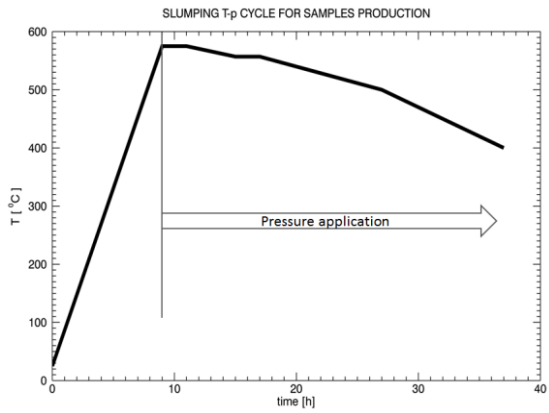

**Fig. 4:** Temperature-pressure cycle used for the realization of all borosilicate D263 samples on Zerodur K20 mould.

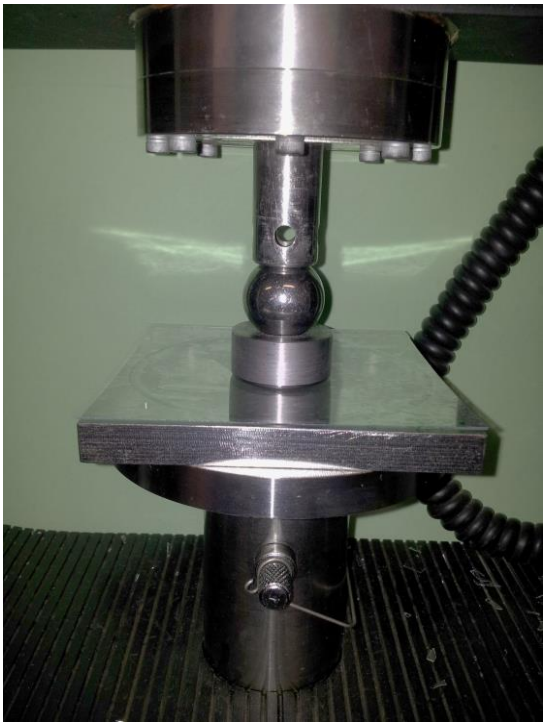

**Fig. 5:** Double ring test configuration used for curved samples. The support and loading rings are clearly visible. The square element surrounding the support ring is used as a template for samples alignment on the machine and does not interfere with the test.

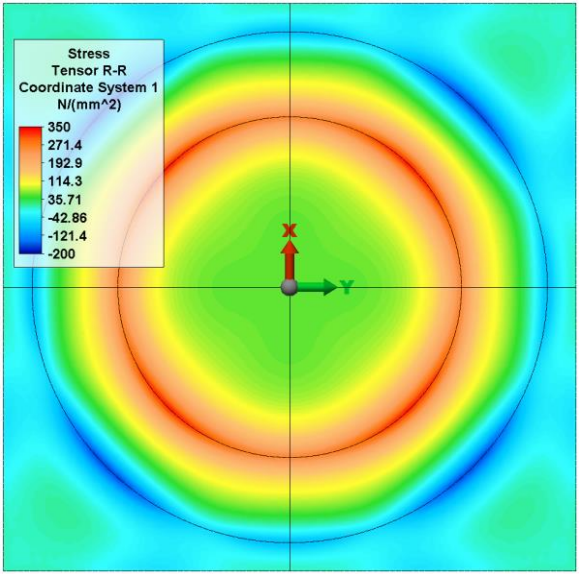

**Fig. 6:** Radial stress field from FEM simulation on the lower surface of flat specimens, beneath a 600 N load; the radial stresses reached their maximum value in correspondence of the loading ring.

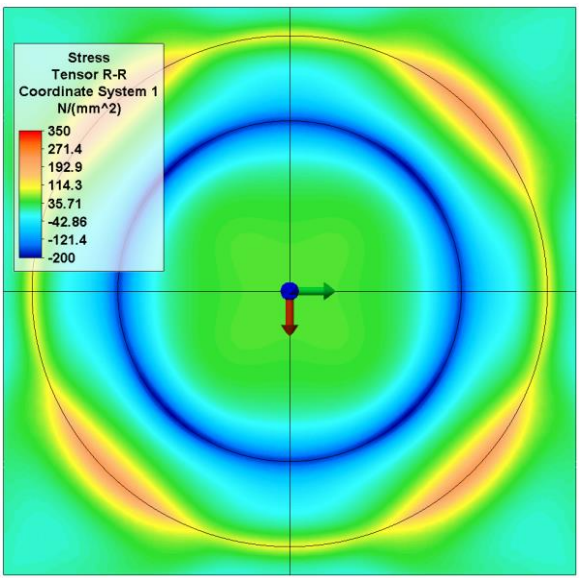

**Fig. 7**: Radial stress field from FEM simulation on the upper surface of flat specimens, beneath a 600 N load; the maximum tensile stress is reached by the radial stress at 45 mm from center, along diagonal.

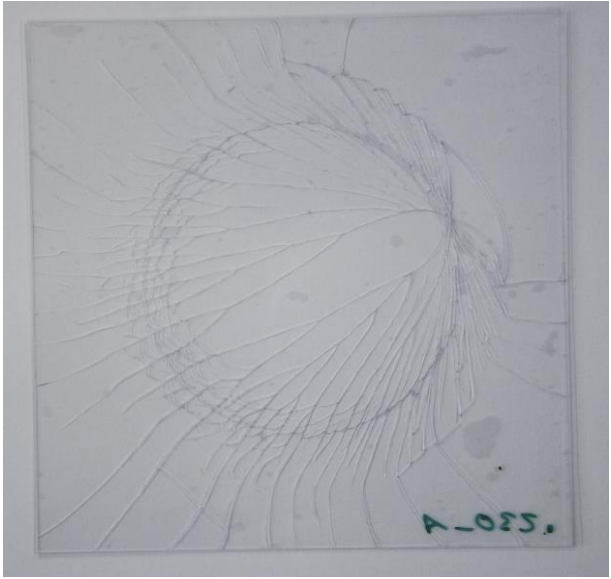

**Fig. 8:** Typical fracture pattern for failures in flat specimens started at the lower (facing-down) surfaces.

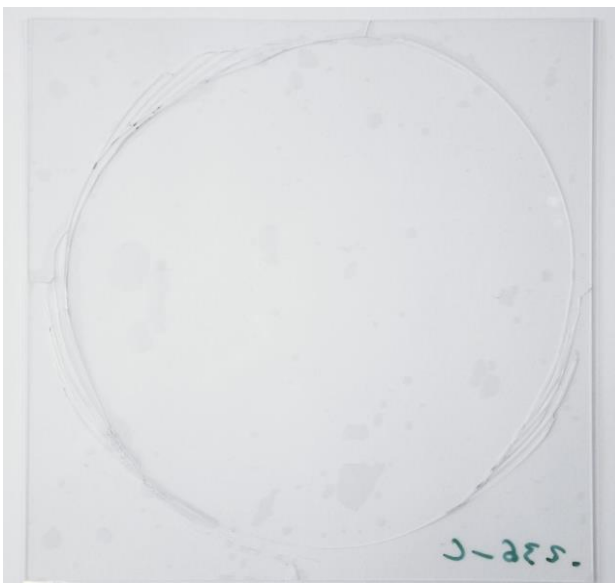

**Fig. 9:** Typical fracture pattern for failures in flat specimens started at the upper (facing-up) surfaces.

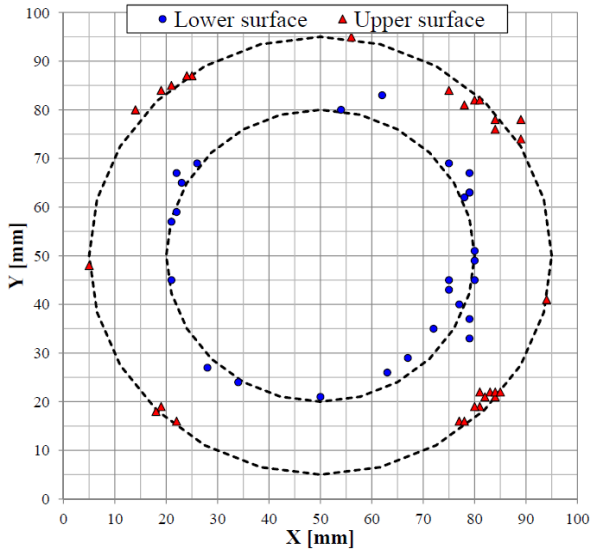

**Fig. 10:** Fracture origins on samples AIR-P (FLAT SPECIMENS). The two circles represent the bearing-ring position (bigger circle) and the loading-ring position (smaller circle).

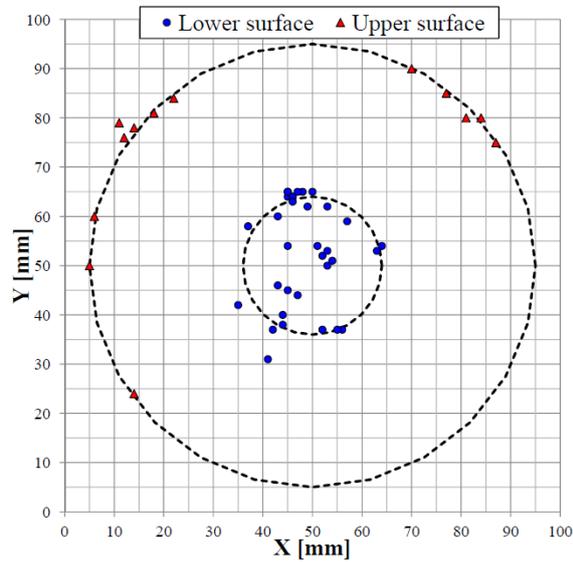

**Fig. 11:** Fracture origins on samples AIR-C (CYLINDRICAL SPECIMENS). The two circles represent the bearing-ring position (bigger circle) and the loading-ring position (smaller circle).

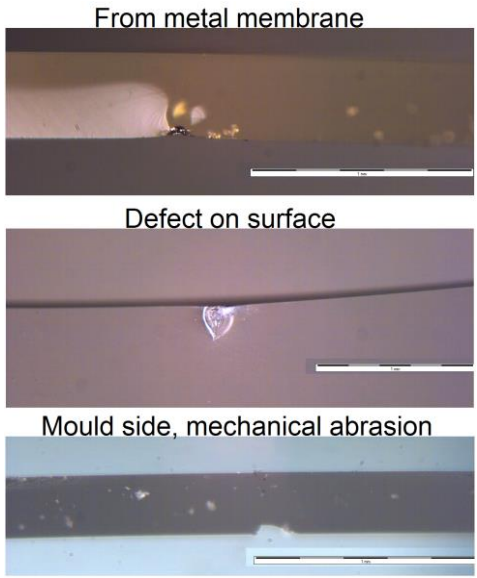

**Fig. 12:** The fractures origin generally from defects on the glass surface, e.g. (Top) defect generated by the contact with the metal pressing membrane; (Center) defect on the back surface probably related to handling issue; (Bottom) mechanical abrasion found on the side of the glass in contact with the mould.

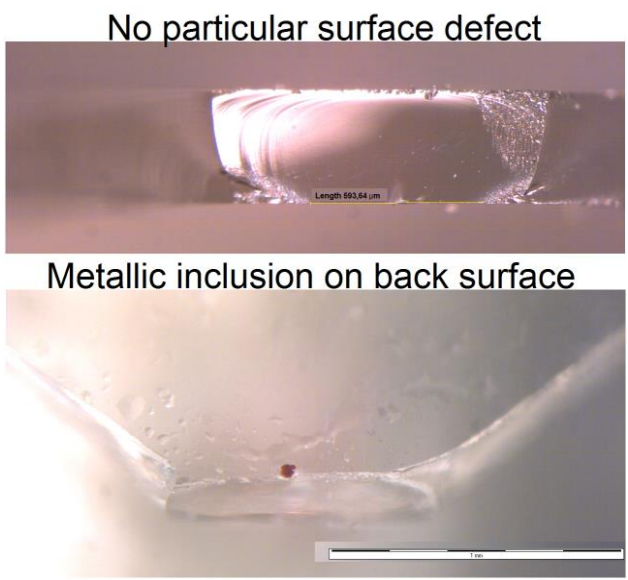

**Fig. 13:** Area of fracture origins for two samples of set AIR-P: (Top) no evident surface defects on the optical surface; (Bottom) metallic inclusion on the back surface, due to non-optimal cleaning of a muffle cover in the oven.

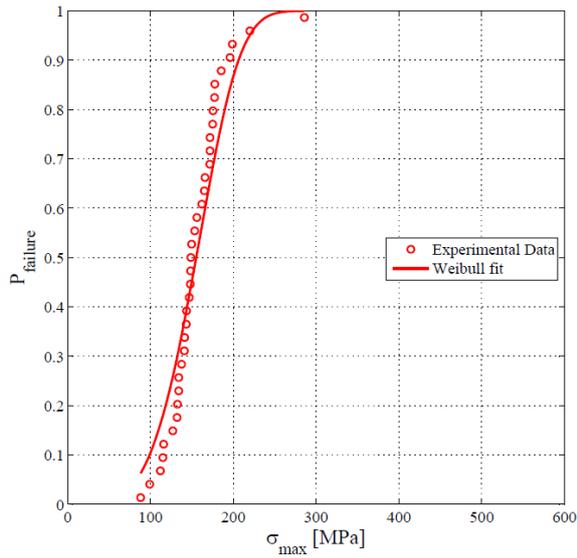

**Fig. 14:** Example of how the Weibull curve fits the experimental data: the particular case refers to MOULD set.

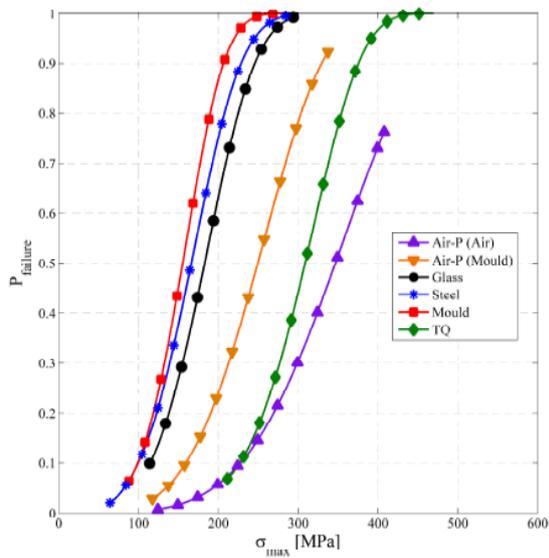

**Fig. 15:** Comparison of the Weibull curves, i.e. fracture probability of the glass plates versus the maximum value of principal tensile stress, obtained for the different sets of flat samples.

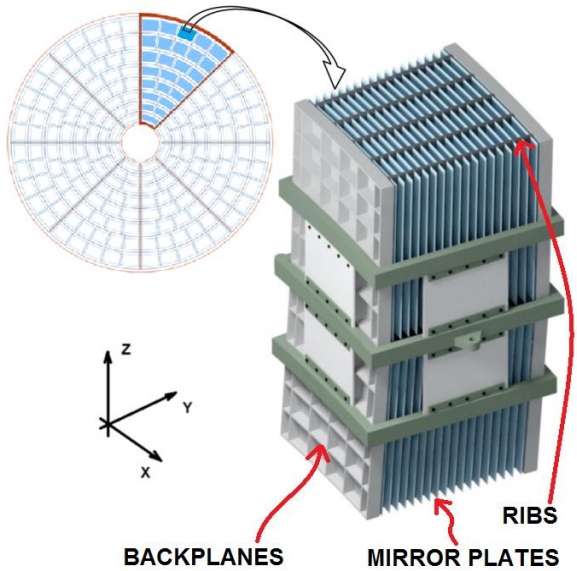

**Fig. 16:** Conceptual design of the flight mirror assembly. The X-ray Optical Units are arranged in 8 ring and 8 petals. The current design of an intermediate XOU is also shown, with indication of the adopted reference system for the considered loading conditions.

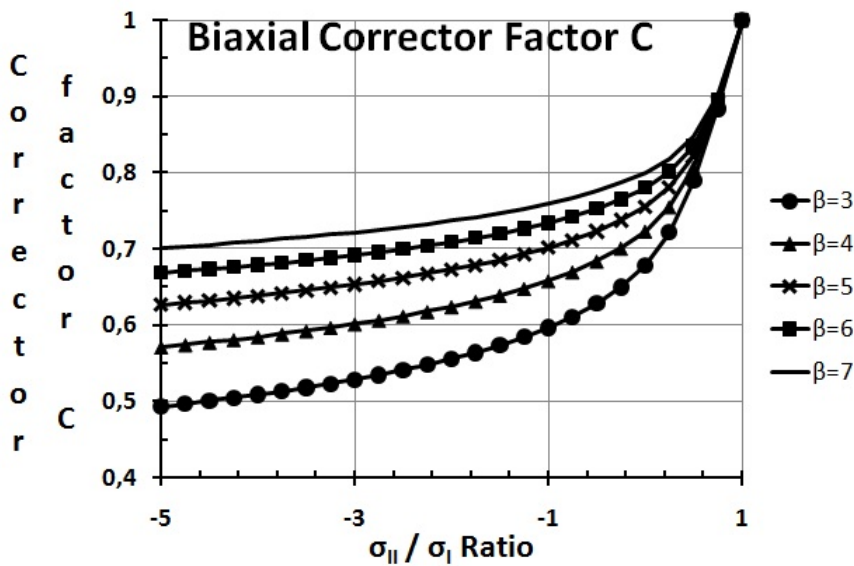

**Fig. 17:** Correction factor taking into account non uniform biaxial stress reported as a function of the ratio between the two principal stresses $\sigma_I$ and $\sigma_{II}$ (where $\sigma_I$ is the maximum tensile stress >0) and in dependence of different values of the Weibull modulus $\beta$.